\documentclass[runningheads]{llncs}
\usepackage[T1]{fontenc}
\usepackage{graphicx}
\begin{document}
\title{Search Plurality}
\author{Shiran Dudy} %\and
%Second Author\inst{2,3}\orcidID{1111-2222-3333-4444} \and
%Third Author\inst{3}\orcidID{2222--3333-4444-5555}}
%
\authorrunning{Shiran Dudy}
\institute{Northeastern University}%,\\ Boston MA 02115, USA}% \and
%Springer Heidelberg, Tiergartenstr. 17, 69121 Heidelberg, Germany
\maketitle              % typeset the header of the contribution
\begin{abstract}
In light of Phillips' contention regarding the impracticality of Search Neutrality~\cite{phillips2023algorithmic}, asserting that non-epistemic factors presently dictate result prioritization, our objective in this study is to confront this constraint by questioning prevailing design practices in search engines. We posit that the concept of prioritization warrants scrutiny, along with the consistent hierarchical ordering that underlies this lack of neutrality.
We introduce the term Search Plurality to encapsulate the idea of emphasizing the various means a query can be approached. This is demonstrated in a design that prioritizes the display of categories over specific search items, helping users grasp the breadth of their search. Whether a query allows for multiple interpretations or invites diverse opinions, the presentation of categories highlights the significance of organizing data based on relevance, importance, and relative significance, akin to traditional methods. However, unlike previous approaches, this method enriches our comprehension of the overall information landscape, countering the potential bias introduced by ranked lists. 

\end{abstract}
\section{Introduction}

A recent work conducted by Phillips~\cite{phillips2023algorithmic} contends that achieving search neutrality within the existing framework is unattainable. According to their interpretation, Search Neutrality denotes a scenario where search outcomes solely reflect relevance and remain uninfluenced by any non-epistemic ideologies. The author further acknowledges the complexity of relevance, highlighting its multi-dimensional nature, which renders these dimensions non-comparable. Despite this non-comparability, search engines still assign priorities to items, suggesting a semblance of comparability, albeit not on a relevance scale. Consequently, such prioritization is susceptible to influence from non-epistemic agendas.

Acknowledging the hurdles in attaining search neutrality, our endeavor in this work is to alleviate the impact of bias perpetuated by these systems and endorse the concept of Search Plurality. Search Plurality revolves around enhancing information access by presenting a diverse array of relevant categories linked to a given query. This approach aims to offer individuals seeking information a broad spectrum of options, aiding in comprehension of the available content, rather than merely navigating through structured lists of items. Such an approach may be particularly beneficial for when search is intended for exploration or learning, as it facilitates a more expansive exploration of available knowledge~\cite{marchionini2006exploratory,white2009exploratory,vakkari2016searching}.

\begin{figure}
\includegraphics[width=\textwidth]{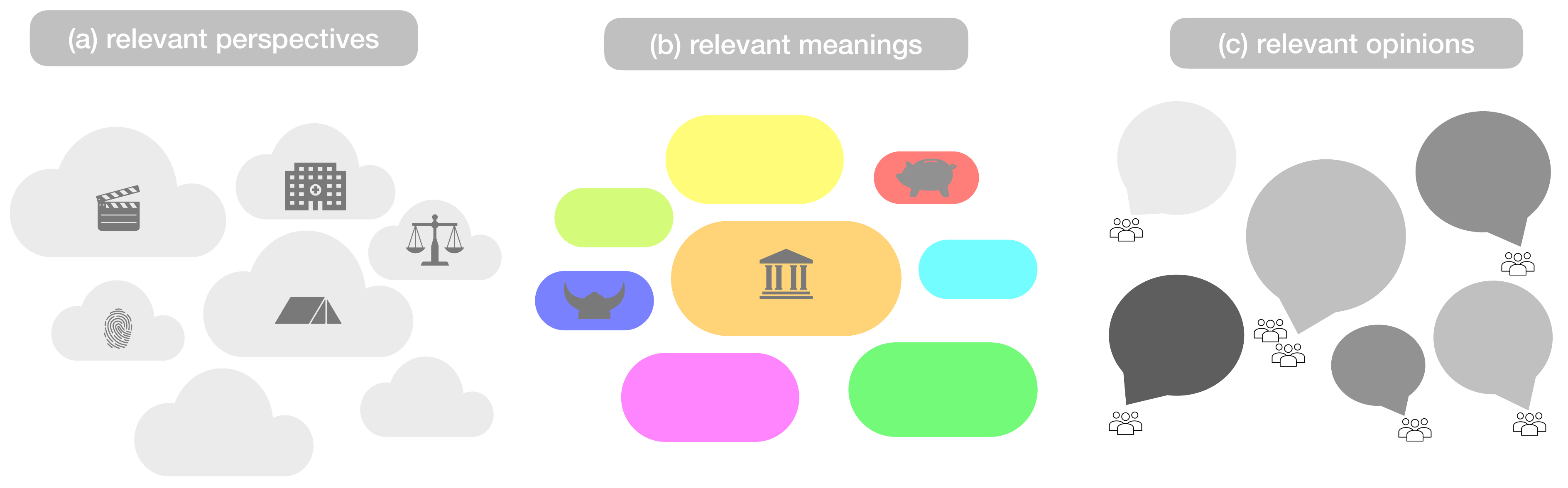}
\caption{Search Plurality is illustrated through three scenarios where the size of the blurb reflects its relative importance to the query (i.e. number of items within a category), and the categories demonstrate the breadth of the query. (a) Presenting a diverse array of perspectives to tackle a query like "how do I avoid being evacuated," encompassing categories pertinent to law, healthcare, media and news, as well as law enforcement. (b) Offering a variety of options to clarify the term and its meaning in a query such as "what is banking it." (c) Providing a spectrum of opinions surrounding a query like "what should I pursue a degree in?"} \label{fig1}
\end{figure}

\section{Search Plurality}
\subsection{Design elements}

In this approach, we propose the incorporation of three elements: \textit{categories}, \textit{explicit relevance}, and \textit{plural design.} A \textit{category} display signifies a specific topic relevant to the query, with its items becoming visible upon zooming in. To cultivate an epistemically oriented environment, transparency regarding the justifications for relevance would be advantageous. Explicit relevance involves indicating a label/icon associated with the blurb.

Furthermore, as outlined in Phillips~\cite{phillips2023algorithmic}, the enforced ranking and hierarchical structuring of items within search results introduces platform bias, which is influenced by factors \textit{other than relevance}. Consequently, this bias has been shown to disproportionately direct users' attention towards items positioned higher in the stack~\cite{chun2022power}. To mitigate users' skewed attention resulting from biased ranking, we can integrate plural design elements that acknowledge and appreciate the multifaceted perspectives inherent in the human experience. This entails presenting labeled blurbs on the screen, where their saliency (or relevance degree) is depicted through their relative size. However, their arrangement on the screen is organized in a manner that does not imply superiority of one category over another due to the incomparability of relevance. This approach would allow users to perceive the depth and breadth of conversations surrounding a particular topic within a single display, rather than grappling with the process of deciphering a domain from an endless list of items.

\subsection{Case studies}

\textbf{Diverse Perspectives} depicts a user who seeks to gain comprehensive knowledge in a domain, where a query can be approached from various angles (illustrated in Figure~\ref{fig1} (a)). For instance, in the scenario outlined in \cite{shah2022situating}, a user asks, "Who can help me avoid being evicted." This query could elicit a range of relevant sources: in the \textit{news} category, information on eviction relief or notable legal cases related to police evictions (assuming geolocation data is available); in the \textit{social activism} category, details about grassroots organizations and their assistance methods; or in the \textit{blogposts} category, discussions across social media platforms on the subject. In the quest for a comprehensive understanding of this topic and potentially taking actionable steps, a variety of sources can provide well-rounded access to relevant resources.

\textbf{Multiple Interpretations} may arise from the query "how can I do AI," (illustrated in Figure~\ref{fig1} (b)) necessitating additional context for clarity. Addressing these ambiguities involves exploring various aspects. Suggestions may include \textit{integrating AI into the workplace}, \textit{enhancing programming skills}, \textit{AI's risks and benefits}, \textit{AI applications}, \textit{DIY AI projects}, etc. Presenting multiple means of contextualizeation demonstrates the multifaceted nature of engaging with this query. By minimizing assumptions, the broad spectrum of responses not only underscores the limited comprehension of the search engine regarding the specific context but also validates the diverse needs of users expressed through their queries. Moreover, it serves as an educational tool, illuminating the various possibilities within the topic.

\textbf{Opinions' Spectrum or Space} can be exemplified by a query such as "who should I vote for in this upcoming election and why?" (illustrated in Figure~\ref{fig1} (c)). This query provides an opportunity to explore a wide range of opinions and understand the rationales behind them, thereby gaining insight into the values and narratives that influence our collective political perspectives. Engaging with viewpoints different from our own fosters tolerance, respect, and empathy towards diverse human experiences and values. Introna and Nissenbaum~\cite{introna2000shaping} already warned that (commercial) search engines are likely to prioritize catering to mainstream interests, often neglecting marginalized groups and communities, and indeed this unfortunate reality was unfolded in \cite{rahman2020algorithms}).\\

\textbf{Limitations:} Similar to a ranked list, this format may not be optimal for \textit{lookup} intent, where users seek factual information or concise answers. In such cases, conversational search engines that offer quick and short responses may be more suitable. Additionally, the introduction of a new design implementation that emphasizes multiple perspectives may obscure direct access to specific items, necessitating a learning curve for users. Furthermore, by promoting diverse viewpoints, it becomes essential to decide which perspectives are included and, consequently, which ones are excluded. It's imperative to ensure that all perspectives, particularly those of underrepresented communities, are given prominence~\cite{rahman2020algorithms}, while viewpoints propagated by conspiracy theorists are not endorsed~\cite{urman2022earth}.

\textbf{Benefits:} The proposed approach benefits learners and explorers by providing us comprehensive support for engaging with a topic, acknowledging the diverse needs we may have, and broadening our understanding of perspectives beyond our own. By doing so, we cultivate epistemic conditions that encourage a more nuanced approach to understand what is relevant, how, and its degree of relevance. It acknowledges that there may not always be a definitive "best" or "worst" answer, as search items are often incomparable but still relevant. Moreover, by prioritizing plurality over hierarchy, we situate ourselves within a context that discourages the formation of echo chambers~\cite{golebiewski2019data} %'s "fragmented concepts"), 
and single narratives~\cite{robertson2018auditing}. This approach ultimately fosters increased compassion and respect for the multitude of valid perspectives and ways of being.

\subsection{Who gets to decide}

The ambitious goal of diversifying search engines involves not only representing a variety of perspectives but also engaging in collective decision-making to determine relevance in participatory ways. Entrusting such decisions to a select few individuals could perpetuate biased ecosystems. Instead, actively participating in this process, akin to the collaborative model of Wikipedia~\cite{kuznetsov2006motivations}, empowers communities to take ownership of their narratives and actions. Through Search Plurality, search engines can serve as mediators, fostering a more inclusive and democratic exchange of information.

\bibliographystyle{splncs04}
\bibliography{ref}
%





% LNCS Homepage, \url{http://www.springer.com/lncs}, last accessed 2023/10/25
% \end{thebibliography}
\end{document}